\documentclass[prd, preprint, nofootinbib, showpacs]{revtex4}

\usepackage{graphicx}
\usepackage{amsfonts}
\usepackage{latexsym}
\usepackage{amssymb}
\usepackage{epsfig}

\begin{document}

\title{
 Higgs portal dark matter in the minimal gauged $U(1)_{B-L}$ model
}

\author{Nobuchika Okada}
 \email{okadan@ua.edu}
 \affiliation{
Department of Physics and Astronomy, 
University of Alabama, Tuscaloosa, AL 35487, USA
}

\author{Osamu Seto}
 \email{seto@phyics.umn.edu}
 \affiliation{
 Department of Architecture and Building Engineering,
 Hokkai-Gakuen University,
 Sapporo 062-8605, Japan
}

%

\begin{abstract}

We propose a scenario of the right-handed neutrino dark matter 
 in the context of the minimal gauged $U(1)_{B-L}$ model 
 by introducing an additional parity which ensures 
 the stability of dark matter particle. 
The annihilation of this right-handed neutrino
 takes place dominantly through the $s$-channel Higgs boson exchange, 
 so that this model can be called Higgs portal dark matter model. 
We show that the thermal relic abundance of 
 the right-handed neutrino dark matter 
 with help of Higgs resonance can match 
 the observed dark matter abundance. 
In addition we estimate the cross section with nucleon  
 and show that the next generation direct dark matter search  
 experiments can explore this model. 

\end{abstract}


\preprint{HGU-CAP 002} 

\vspace*{3cm}
\maketitle


\section{Introduction}

The nonvanishing neutrino masses have been confirmed 
 by various neutrino oscillation phenomena and indicate 
 the evidence of new physics beyond the Standard Model. 
The most attractive idea to naturally explain the tiny 
 neutrino masses is the seesaw mechanism \cite{Seesaw},  
 in which the right-handed (RH) neutrinos singlet under 
 the SM gauge group are introduced. 
The minimal gauged $U(1)_{B-L}$ model based on the 
 gauge group $SU(3)_C \times SU(2)_L \times U(1)_Y \times U(1)_{B-L}$ 
 \cite{Mohapatra:1980qe} is an elegant and simple
 extension of the SM, in which the RH neutrinos 
 of three generations are necessarily introduced 
 because of the gauge and gravitational anomaly cancellations. 
In addition, the mass of RH neutrinos arises 
 associated with the $U(1)_{B-L}$ gauge symmetry breaking.

Although the scale of the $B-L$ gauge symmetry breaking 
 is basically arbitrary as long as phenomenological 
 constraints are satisfied, one interesting option is 
 to take it to be the TeV scale \cite{Khalil:2006yi}. 
It has been recently pointed out \cite{Iso:2009ss} 
 that when the classical conformal invariance is 
 imposed on the minimal $U(1)_{B-L}$ model, 
 the symmetry breaking scale appears to be the TeV scale 
 naturally. 
If this is the case, all new particles, the $Z'$ gauge boson, 
 the $B-L$ Higgs boson $H$ and the RH neutrinos 
 appear at the TeV scale 
 unless the $U(1)_{B-L}$ gauge coupling is extremely small, 
 and they can be discovered at 
 Large Hadron Collider~\cite{Emam:2007dy,Huitu:2008gf,Basso:2008iv,Perez:2009mu}.
Then we may be able to understand the relation 
 between the gauge symmetry breaking and the origin of neutrino masses.

Although such a TeV scale model is interesting and appealing, 
 one might think that the absence of dark matter (DM) candidate 
 is a shortcoming of this model. 
A sterile RH neutrino with mass of the order of MeV 
 is one possibility~\cite{Khalil:2008kp}. 
In this paper, we propose a very simple idea to introduce 
 the DM candidate in the minimal gauged $U(1)_{B-L}$ model. 
We introduce the $Z_2$ parity into the model and impose 
 one of three RH neutrinos to be odd, 
 while the others even. 
In this way, the $Z_2$-odd RH neutrino becomes stable 
 and the DM candidate. 
Note that two RH neutrinos are enough to reconcile with 
 the observed neutrino oscillation data, 
 with a prediction of one massless light neutrino. 
Therefore, without introducing any additional new dynamical 
 degrees of freedom, the DM particle arises 
 in the minimal gauged $U(1)_{B-L}$ model. 

The paper is organized as follows. 
In the next section, we briefly describe our model. 
In section~III, we estimate the thermal relic density of 
 the RH neutrino and identify the model parameter 
 to be consistent with the current observations. 
We also calculate the scattering cross section between 
 the DM particle and nucleon 
 and discuss the implication for the direct DM 
 search experiments. 
We summarize our results in the section~IV. 
Our notations and the formulas used in our analysis 
 are listed in Appendix.

\section{The minimal gauged $U(1)_{B-L}$ model with $Z_2$ parity} 
The model is based on the gauge group 
 $SU(3)_C \times SU(2)_L \times U(1)_Y \times U(1)_{B-L}$. 
Additional fields besides the standard model fields are
 a gauge field $Z^\prime_{\mu}$ of the $U(1)_{B-L}$, 
 a SM singlet $B-L$ Higgs boson $\Psi$ with 
 two $U(1)_{B-L}$ charge, and three RH neutrinos $N_i$ 
 which are necessary for the gauge and gravitational anomaly cancellations.  
In describing the RH neutrinos, we use 
 the four component representation of RH neutrino 
 constructed from the Weyl spinor $\nu_{R_i}$, 
\begin{equation}
 N_i \equiv \left( 
            \begin{array}{c}
            \nu_{R_i} \\
            \epsilon \, \nu_{R_i}^*
            \end{array}
              \right),
\end{equation}
For the two RH neutrinos, $N_1$ and $N_2$, we assign 
 $Z_2$ parity even, while odd for $N_3$, so that
 the RH neutrino $N_3$ is stable and, hence, the DM candidate.

Due to the additional gauge symmetry $U(1)_{B-L}$, 
 the covariant derivative for each fields is given by 
\begin{equation}
 D_{\mu}= D_{\mu}^{(SM)} -i q_{B-L}g_{B-L}Z^\prime_{\mu},
\end{equation}
 where $D_{\mu}^{(SM)}$ is the covariant derivative in the SM, 
 and $q_{B-L}$ is the charge of each fields under the $U(1)_{B-L}$
 with its gauge coupling $g_{B-L}$.

Yukawa interactions relevant for the neutrino masses are given by
\begin{equation}
 {\cal L}_{int} = \sum_{\alpha=1}^3 \sum_{i=1}^2 
   y_{\alpha i} \bar{L}_\alpha \tilde{\Phi} N_i  
   - \frac{1}{2} \sum_{i=1}^3 \lambda_{R_i} 
     \bar{N}_i \Psi P_R N_i + {\rm h.c.},
\end{equation}
 where $\tilde{\Phi} = -i \tau_2 \Phi^*$ for $\Phi$ 
 being the SM Higgs doublet, and without loss of generality 
 we have worked out in the basis where 
 the second term in the right-hand-side is in flavor diagonal 
 for RH neutrinos. 
Because of the $Z_2$ parity, the DM candidate  
 $N_3$ has no Yukawa couplings with the left-handed lepton doublets.

The general Higgs potential for the $SU(2)_L$ doublet $\Phi$ and 
 a singlet $B-L$ Higgs $\Psi$ is generally given by  
\begin{eqnarray}
 V(\Phi, \Psi) = m_1^2 |\Phi|^2 + m_2^2 |\Psi|^2 +\lambda_1 |\Phi|^4 
  + \lambda_2|\Psi|^4 + \lambda_3 |\Phi|^2|\Psi|^2 .
 \label{HiggsPot}
\end{eqnarray}
The Higgs fields $\phi$ and $\psi$ are obtained 
 by expanding $\Phi$ and $\Psi$ as
\begin{eqnarray}
 \Phi &=& \left( 
            \begin{array}{c}
            0 \\
            \frac{1}{\sqrt{2}}(v + \phi) 
            \end{array}
              \right) , \\
 \Psi &=& \frac{1}{\sqrt{2}}(v' + \psi) ,
\end{eqnarray}
 around the true vacuum with the vacuum expectation values $v$ and $v'$.
These are related with the mass eigenstates $h$ and $H$ through
\begin{eqnarray}
 \left( 
            \begin{array}{c}
            h \\
            H 
            \end{array}
 \right)
   &=&
 \left( 
            \begin{array}{cc}
            \cos\theta & -\sin\theta  \\
            \sin\theta & \cos\theta
            \end{array}
 \right) \left( 
            \begin{array}{c}
            \phi \\
            \psi 
            \end{array}
  \right) ,
 \label{HiggsRotation}
\end{eqnarray}
 with $\theta$ being the mixing angle.
Their masses are given by
\begin{eqnarray}
M_h^2 &=& 2\lambda_1 v^2 \cos^2\theta + 2\lambda_2 v'^2 \sin^2\theta 
          - 2\lambda_3 v v' \sin\theta\cos\theta , \label{Mh} \\
M_H^2 &=& 2\lambda_1 v^2 \sin^2\theta + 2\lambda_2 v'^2 \cos^2\theta 
          + 2\lambda_3 v v' \sin\theta\cos\theta . \label{MH}
\end{eqnarray}

The mass of the new neutral gauge boson $Z'$ 
 arises by the $U(1)_{B-L}$ gauge symmetry breaking, 
\begin{equation}
 M_{Z'}^2 = 4 g_{B-L}^2 v'^2. 
\end{equation}
Associated with the $U(1)_{B-L}$ gauge symmetry breaking, 
 the RH neutrinos $N_i$ acquire masses 
\begin{equation}
 M_{N_i} = -\lambda_{R_i} \frac{v'}{\sqrt{2}}.
\end{equation}
 From LEP experiment, the current lower bound 
 on the $Z'$ boson mass has been found to 
 be~\cite{LEPbound1, LEPbound2} 
\begin{equation}
 \frac{M_{Z'}}{g_{B-L}} = 2 v' \gtrsim 6-7 \; {\rm TeV}. 
 \label{BLbound}
\end{equation}
Two $Z_2$-even RH neutrinos $N_1$ and $N_2$ are responsible 
 for light neutrino masses via the seesaw mechanism, 
\begin{equation}
 m_{\nu_{\alpha\beta}}
   = - \sum_{i=1,2} y_{\alpha i}y_{i\beta} \frac{v^2}{2 M_{N_i}} .
\end{equation}
Note that the rank of this mass matrix is two, so that 
 the lightest neutrino is massless.

\section{Right-handed neutrino dark matter}

Due to the $Z_2$ parity, one of RH neutrino $N_3$  
 (we denote it as $N$ hereafter) in our model can be 
 the DM candidate. 
We first estimate its relic abundance and identify 
 the model parameters to be consistent with 
 the current observations. 
Next we calculate the scattering cross section between  
 the DM particle and a proton 
 and discuss the implication for the direct 
 DM search experiments.

\subsection{Thermal relic density}

The DM RH neutrino interacts with the SM particles through
 couplings with $B-L$ gauge and $B-L$ Higgs bosons.
Note that neutrino Dirac Yukawa interactions are absent 
 because of the $Z_2$ parity.
The most of annihilation of the RH neutrinos occurs 
 via $Z', H$ and $h$ exchange processes in the $s$-channel. 
In practice, the dominant contributions come from 
 the Higgs ($h$ and $H$) exchange diagrams, 
 because the $Z'$ exchange processes are suppressed by 
 the inverse square of the $B-L$ Higgs VEV $v' \gtrsim 3$ TeV. 
Thus, we obtain Higgs portal DM of RH neutrino effectively. 
The relevant annihilation modes are the annihilation
 into $f \bar{f}$, $W^+ W^-$, $Z Z$, and $h(H) h(H)$.
Since RH neutrino DM couples to only $B-L$ Higgs $\Psi$ 
 while a SM particle does to SM Higgs $\Phi$,
 the DM annihilation occurs only through the mixing 
 between these two Higgs bosons. 
Although it is not so severe, 
 the precision electroweak measurements~\cite{Dawson:2009yx} 
 as well as the unitarity bound~\cite{Basso:2010jt}
 give constraints on the mixing angle and mass spectrum of the Higgs bosons. 

The thermal relic abundance of DM
\begin{equation}
\Omega_N h^2 =
 1.1 \times 10^9 \frac{m_N/T_d}{\sqrt{g_*}M_P\langle\sigma v\rangle}
  {\rm GeV^{-1}} ,
\end{equation}
 with the Planck mass $M_P$, 
 the thermal averaged product of the annihilation cross section 
 and the relative velocity $\langle\sigma v\rangle$,
 the total number of relativistic degrees of freedom in the thermal bath $g_*$, 
 and the decoupling temperature $T_d$, is evaluated by solving 
 the Boltzmann equation for the number density of RH neutrino $n_N$;
\begin{equation}
 \frac{d n_N}{dt}+3H n_N=-\langle\sigma v\rangle(n_N^2 - n_{\rm EQ}^2),
\end{equation}
 and the Friedmann equation
\begin{eqnarray}
 H^2 \equiv \left( \frac{\dot{a}}{a} \right)^2 = \frac{8 \pi }{3M_P^2} \rho ,
\end{eqnarray}
 with $n_{\rm EQ}$ and $a(t)$ being the equilibrium number density
 and the scale factor, 
 under the radiation dominated Universe with 
 the energy density $\rho=\rho_{\rm rad}$~\cite{O56:KolbTurner}.

Fig.~1 shows the relic density $\Omega_N h^2$ as a function of
 the DM mass $m_N$ for a set of parameters: 
 $(v', M_h, M_H, M_{Z'}, \sin\theta) 
 = (4000 \;{\rm GeV}, 120 \; {\rm GeV}, 200 \; {\rm GeV}, 1000 \;{\rm
 GeV}, 0.7)$, for example. 
Willkinson Microwave Anisotropy Probe measured the value of DM
 abundance as $\Omega_{DM}h^2 \simeq 0.1$~\cite{WMAP}.
The figure shows that a desired DM relic abundance 
 can be obtained for only near Higgs resonances, 
 $m_N \approx M_h/2$ or $M_H/2$.

\begin{figure}[h,t]
\begin{center}
\epsfig{file=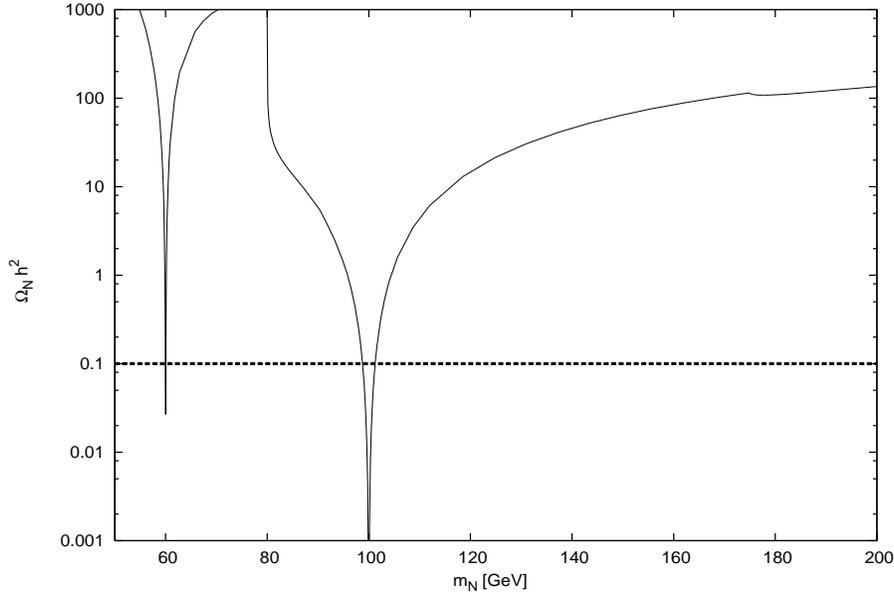, width=12cm,height=8cm,angle=0}
\end{center}
\caption{
 The thermal relic density of RH neutrino DM 
 as a function of its mass for a parameter set: 
 $(v', M_h, M_H, M_{Z'}, \sin\theta) 
 = (3000 \; {\rm GeV}, 120 \; {\rm GeV}, 200 \; {\rm GeV}, 1000 \;{\rm GeV}, 0.7)$.
 }
\label{fig1}
\end{figure}

Fig.~2 shows the relic density $\Omega_N h^2$ as a function of
 the DM mass $m_N$ for a smaller Higgs mixing $\sin\theta = 0.3$ 
 (others are the same as in Fig.~1). 
Compared with Fig.~\ref{fig1}, 
 for $m_N \lesssim M_W$ where the DM particles 
 dominantly annihilate into $f \bar{f}$, 
 the relic density further increases 
 because of the small mixing angle. 
When the DM is heavier, the annihilation mode into Higgs boson pairs 
 is opened and the relic density slightly deceases, 
 but the reduction is not enough to reach 
 $\Omega_N h^2 \simeq 0.1$.

%

\begin{figure}[h,t]
\begin{center}
\epsfig{file=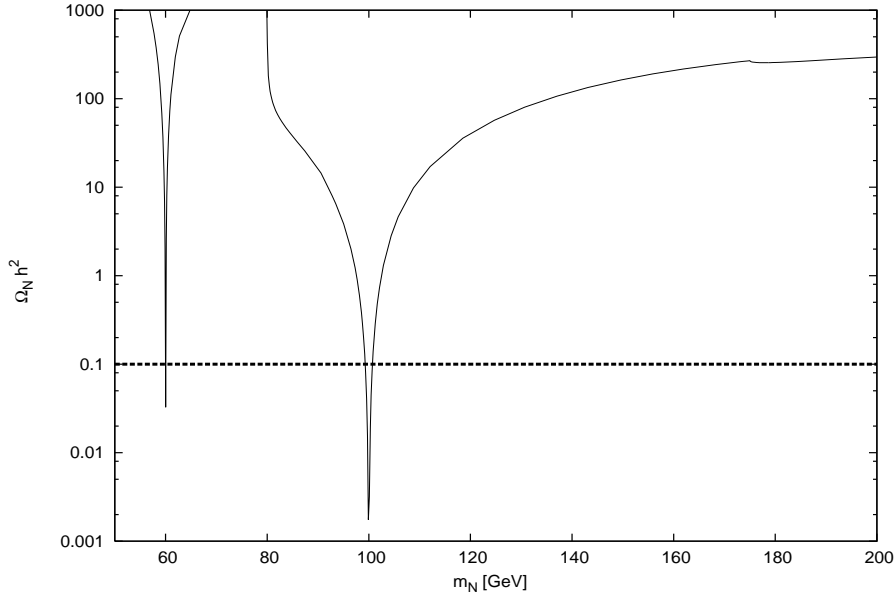, width=12cm,height=8cm,angle=0}
\end{center}
\caption{
 The same as Fig.~\ref{fig1} but for $\sin \theta = 0.3$.
 }
\label{fig2}
\end{figure}

Our model is quite analogous to the so-called gauge singlet scalar dark
 matter~\cite{McDonald:1993ex,Burgess:2000yq,Davoudiasl:2004be}.
Some recent studies can be found in Refs.~\cite{Kikuchi:2007az, Yaguna:2008hd}.
In the gauge singlet scalar DM model,
 the thermal abundance is mainly controlled 
 by the interactions between 
 the SM Higgs boson and the DM particle. 
In our model, $B-L$ Higgs VEV $v'$ can play the same role for $m_N < M_W$,
 namely a larger $v'$ corresponds to weaker coupling between DM and Higgs
 for a fixed DM mass.
On the other hand, for $m_N > M_W$ the difference appears.
Even if the annihilation mode into $W$-boson pair becomes kinematically available,
 it is not possible to obtain the desired DM abundance 
 without the Higgs resonant annihilation
 because the bound on $v'$ given by Eq.~(\ref{BLbound}) is stringent.

\subsection{Direct detection of dark matter} 

Our RH neutrino DM can elastically scatter off with nucleon,  
 unlike another RH neutrino DM model 
 has been proposed by Krauss {\it et. al.}~\cite{Krauss:2002px} 
 and studied~\cite{Baltz:2002we,Cheung:2004xm}.
The main process is Higgs exchange and
 the resultant cross section for a proton is given by
\begin{equation}
\sigma_{\rm SI}^{(p)} = \frac{4}{\pi}
 \left(\frac{m_p m_N}{m_p + m_N}\right)^2 f_p^2, 
 \label{sigmaSI}
\end{equation}
 with the hadronic matrix element
\begin{equation}
 \frac{f_p}{m_p} = \sum_{q=u,d,s}f_{Tq}^{(p)}\frac{\alpha_q}{m_q} 
  + \frac{2}{27}f_{TG}^{(p)}\sum_{c,b,t}\frac{\alpha_q}{m_q},
\end{equation}
 and the effective vertex (see Appendix for notations) 
\begin{equation}
 \alpha_q = -\lambda_N y_q  \left(\frac{\partial \Phi}{\partial h}
 \frac{1}{M_h^2 } \frac{\partial \Psi}{\partial h} 
 + \frac{\partial \Phi}{\partial H} \frac{1}{M_H^2}
 \frac{\partial \Psi}{\partial H} \right) ,
 \label{alphaq}
\end{equation}
 where $m_q$ is a mass of a quark with a Yukawa coupling $y_q$, and
  $f_{Tq}^{(p)}$ and $f_{TG}^{(p)}$ are constants.

From Eq.~(\ref{alphaq}), 
 one can see that $\sigma_{\rm SI}^{(p)} \propto (\sin2\theta/v')^2$ 
  for a given DM mass $m_N$. 
Fig.~3 shows the spin-independent cross section of RH neutrino 
 with a proton. 
The resultant cross section is found to be far below 
 the current limits reported by 
 XENON10~\cite{XENON10} and CDMSII~\cite{CDMSII}: 
 $ \sigma_{\rm SI}
 \lesssim 4 \times 10^{-8} -  2 \times 10^{-7} $ pb, 
 for a DM mass of 100 GeV-1 TeV. 
Future experiments such as XENON1T~\cite{XENON100} 
 can reach the cross section predicted in our model.

\begin{figure}[h,t]
\begin{center}
\epsfig{file=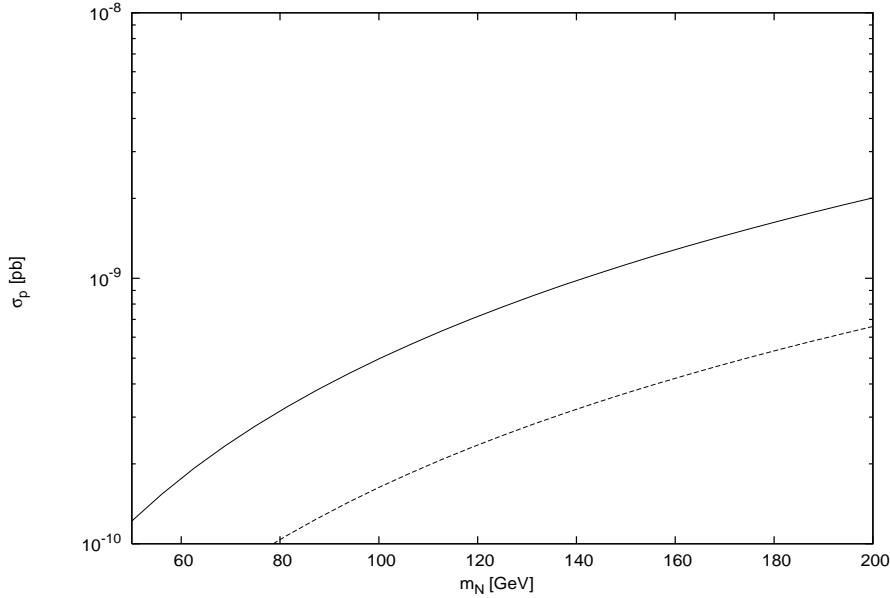, width=12cm,height=8cm,angle=0}
\end{center}
\caption{
 The spin independent scattering cross section with a proton.
 All parameters are same as those used in the previous section.
 The upper and lower lines correspond to $\sin\theta = 0.7$ and $0.3$, 
 respectively. 
 }
\label{fig3}
\end{figure}
%

\section{Summary}

We have proposed a scenario of the RH neutrino dark matter 
 in the context of the minimal gauged $U(1)_{B-L}$ model. 
We have introduced a discrete $Z_2$ parity in the model, 
 so that one RH neutrino assigned as $Z_2$-odd 
 can be stable and, hence, the DM candidate, 
 while the other two RH neutrinos account 
 for neutrino masses and mixings through the seesaw mechanism. 
No additional degrees of freedom are necessary to be added. 
We have evaluated the relic density of the dark matter particle. 
The dominant annihilation modes are via the Higgs boson exchange 
 processes in the $s$-channel and thus, our model can be called 
 Higgs portal DM model. 
It has been found that the relic density consistent with 
 the current observation can be achieved 
 only when the annihilation processes are enhanced 
 by Higgs resonances. 
Therefore, the mass of the RH neutrino DM 
 should be around a half of Higgs boson masses. 
We have also calculated the elastic scattering cross section 
 between the DM particle and a proton 
 and found it within the reach of future experiments 
 for the direct DM search.

%
%
%

\appendix

\section{The Higgs sector} 

The Higgs potential (\ref{HiggsPot}) contains five parameters:
 $m_1^2, m_2^2, \lambda_1, \lambda_2$ and $\lambda_3$.
These parameters can be rewritten in terms of two Higgs VEVs, 
 two physical Higgs masses and the mixing angle between them.
The stationary conditions are
\begin{eqnarray}
&& m_1^2 + \lambda_1 v^2 + \frac{1}{2}\lambda_3 v'^2 = 0 , \\
&& m_2^2 + \lambda_2 v^2 + \frac{1}{2}\lambda_3 v'^2 = 0 .
\end{eqnarray}
The physical Higgs masses are given by Eqs.~(\ref{Mh}) and (\ref{MH}) with 
 the mixing angle that $\theta$ satisfies 
\begin{equation}
 \tan2\theta
    = - \frac{\lambda_3 v v'}{(\lambda_1 v^2 -\lambda_2 v'^2)} .
\end{equation}

Higgs self interaction terms are expressed as
\begin{equation}
 {\cal L}_{int}
    = \lambda_1 v \phi^3 +\lambda_2 v' \psi^3 
     + \frac{1}{2}\lambda_3(v \phi \psi^2 + v' \psi \phi^2)
     + \frac{1}{4}(\lambda_1 \phi^4 + \lambda_2 \psi^4 +\lambda_3 \phi^2\psi^2), 
\end{equation}
 in terms of $\phi$ and $\psi$. With Eq.~(\ref{HiggsRotation}),
 these are rewritten in terms of $h$ and $H$ with $\theta$ as
\begin{eqnarray}
&& {\cal L}_{int} \nonumber \\
 &=&
  \left[\lambda_1 v \cos^3\theta - \lambda_2 v' \sin^3\theta + \frac{1}{2}\lambda_3(v \cos\theta \sin^2\theta - v' \sin\theta \cos^2\theta)\right] hhh \nonumber \\
  && +\left[ 3 \lambda_1 v \cos^2\theta\sin\theta 
  + 3 \lambda_2 v' \sin^2\theta\cos\theta   + \frac{1}{2}\lambda_3(v (\sin^3\theta - 2 \cos^2\theta\sin\theta ) 
 \right. \nonumber \\ && \left. + v' (\cos^3\theta-2 \sin^2\theta\cos\theta)) \right] hhH \nonumber \\
  && + \left[3 \lambda_1 v \cos\theta\sin^2\theta -3 \lambda_2 v' \sin\theta\cos^2\theta + \frac{1}{2}\lambda_3(v (\cos^3\theta-2\sin^2\theta\cos\theta)  
  \right. \nonumber \\ && \left. + v' (-\sin^3\theta+2\sin\theta\cos^2\theta))\right] hHH \nonumber \\
  && + \left[\lambda_1 v \sin^3\theta +\lambda_2 v' \cos^3\theta + \frac{1}{2}\lambda_3(v \sin\theta \cos^2\theta + v' \sin^2\theta\cos\theta)\right] HHH \nonumber \\
  && + {\rm four \,~\, point \,~\, interactions}.
 \label{HiggshHselfcoupling}
\end{eqnarray}
We can read off a Higgs three point vertex from Eq.~(\ref{HiggshHselfcoupling}).

In the expression of annihilation cross section,
 we used the following notations :
\begin{eqnarray}
&& \frac{\partial \Phi}{\partial h}= \frac{1}{\sqrt{2}}\cos\theta, \nonumber \\
&& \frac{\partial \Phi}{\partial H}= \frac{1}{\sqrt{2}}\sin\theta, \nonumber \\
&& \frac{\partial \Psi}{\partial h}= -\frac{1}{\sqrt{2}}\sin\theta, \nonumber \\
&& \frac{\partial \Psi}{\partial H}= \frac{1}{\sqrt{2}}\cos\theta.
\end{eqnarray}
%

\section{Amplitude}

%

We give explicit formulas of the invariant amplitude 
 squared for the pair annihilation processes 
 of the RH neutrinos. 
 
\subsection{Annihilation into charged fermions}


%
\begin{eqnarray}
&& \left|{\cal M}\right|^2 =  \nonumber \\
&& 32\left|\frac{g_{B-L}^2 q_f q_N}{s-M_{Z'}^2+i M_{Z'}\Gamma_{Z'}}\right|^2
 (s-4 m_N^2)\left(  
  \frac{3}{8} s -\frac{1}{2}\left(\frac{s}{2}-m_f^2 \right)
  +\frac{1}{2}\left(\frac{s}{4}-m_f^2\right)\cos^2\theta 
  \right)  \nonumber \\
&& 
 + 16 \lambda_N^2
 \left|y_f \left(\frac{\partial \Phi}{\partial h}
 \frac{i}{s - M_h^2 + i M_h \Gamma_h }
 \frac{\partial \Psi}{\partial h} 
 + \frac{\partial \Phi}{\partial H}
 \frac{i}{s - M_H^2 + i M_H \Gamma_H }
 \frac{\partial \Psi}{\partial H}  \right) \right|^2  \nonumber \\
&& 
 (s-4 m_N^2)\left(\frac{s}{4}-m_f^2\right) .
\end{eqnarray}
%

\subsection{Annihilation into neutrinos}
\subsubsection{Annihilation into $\nu_a, \nu_a$ (light active-like neutrinos)}
\begin{eqnarray}
&& \left|{\cal M}\right|^2 =  \nonumber \\
&& 32\left|\frac{g_{B-L}^2 q_f q_N}{s-M_{Z'}^2+i M_{Z'}\Gamma_{Z'}}\right|^2
 (s-4 m_N^2)\left(  
  \frac{3}{8} s -\frac{1}{2}\left(\frac{s}{2}+m_{\nu_a}^2 \right)
  +\frac{1}{2}\left(\frac{s}{4}+m_{\nu_a}^2\right)\cos^2\theta 
  \right)  .
\end{eqnarray}

\subsubsection{Annihilation into $\nu_s, \nu_s$ (heavy sterile-like neutrinos)}
\begin{eqnarray}
&& \left|{\cal M}\right|^2 =  \nonumber \\
&& 32\left|\frac{g_{B-L}^2 q_f q_N}{s-M_{Z'}^2+i M_{Z'}\Gamma_{Z'}}\right|^2
 (s-4 m_N^2)\left(  
  \frac{3}{8} s -\frac{1}{2}\left(\frac{s}{2}+m_{\nu_s}^2 \right)
  +\frac{1}{2}\left(\frac{s}{4}+m_{\nu_s}^2\right)\cos^2\theta 
  \right)  \nonumber \\
&&  + 4 \lambda_N^2 \lambda_{\nu_s}^2 
 \left|\frac{\partial \Psi}{\partial h}
 \frac{i}{s - M_h^2 + i M_h \Gamma_h }
 \frac{\partial \Psi}{\partial h} 
 + \frac{\partial \Psi}{\partial H}
 \frac{i}{s - M_H^2 + i M_H \Gamma_H }
 \frac{\partial \Psi}{\partial H}  \right|^2  
 (s-4 m_N^2)(s-4 m_{\nu_s}^2) . \nonumber \\
  && 
\end{eqnarray}
\subsection{Annihilation into $W^+ W^-$}

\begin{eqnarray}
|{\cal M}|^2
 &=& 8\lambda_N^2 \left(\frac{1}{2}g^2 v \right)^2 
  \left|\frac{\partial \Psi}{\partial h}
  \frac{1}{s-M_h^2+i M_h \Gamma_h} \frac{\partial \phi}{\partial h}
  + \frac{\partial \Psi}{\partial H}
  \frac{1}{s-M_H^2+i M_H \Gamma_H} \frac{\partial \phi}{\partial H} \right|^2  
  \nonumber \\
  && (s-4m_N^2) \left( 1 +\frac{1}{2 M_W^4}\left(\frac{s}{2} - M_W^2\right)^2 \right) .
\end{eqnarray}
\subsection{Annihilation into $Z Z$}

\begin{eqnarray}
|{\cal M}|^2
 &=& 8\lambda_N^2 \left(\frac{1}{4}(g^2+g'^2 )v \right)^2 
  \left|\frac{\partial \Psi}{\partial h}
  \frac{1}{s-M_h^2+i M_h \Gamma_h} \frac{\partial \phi}{\partial h}
  + \frac{\partial \Psi}{\partial H}
  \frac{1}{s-M_H^2+i M_H \Gamma_H} \frac{\partial \phi}{\partial H} \right|^2  
  \nonumber \\
  && 
  (s-4m_N^2) \left( 1 +\frac{1}{2 M_Z^4}\left(\frac{s}{2} - M_Z^2\right)^2 \right) .
\end{eqnarray}
\subsection{Annihilation into $h h$}

${\cal M}_1 $ denotes the amplitude by $s$-channel Higgs bosons $h$ and $H$ exchange,
 while ${\cal M}_2 $ does that for $t(u)$-channel $N$ exchange diagram.
The formulas for $N N \rightarrow h H$ and $H H$ can be obtained by 
 appropriate replacement of the vertexes, e.g., $\lambda_{hhh} \rightarrow\lambda_{hhH}$.
\begin{eqnarray}
|{\cal M}|^2 &=& |{\cal M}_1+{\cal M}_2|^2 , \\
|{\cal M}_1|^2
 &=&
  \lambda_N^2 \left(\frac{s}{2}-2 m_N^2 \right) \nonumber \\
&&
  \left|\frac{\partial \Psi}{\partial h}
  \frac{i}{s-M_h^2+i M_h \Gamma_h} i \lambda_{hhh}
  + \frac{\partial \Psi}{\partial H}
  \frac{i}{s-M_H^2+i M_H \Gamma_H} i \lambda_{Hhh} \right|^2 , \\
\int\frac{d \cos\theta}{2}|{\cal M}_2|^2
 &=&
  \lambda_N^4 \left( \frac{\partial \Psi}{\partial h}\right)^4
  \left(
   -8 - I_{22} + J_{22} \ln\left|\frac{A+2 b}{A-2 b}\right|
  \right) , \\
\int\frac{d \cos\theta}{2} {\cal M}_1 {\cal M}_2^*
 &=&
  4 m_N \lambda_N^3 \left( \frac{\partial \Psi}{\partial h}\right)^2
  \left( \frac{\partial \Psi}{\partial h}
  \frac{i}{s-M_h^2+i M_h \Gamma_h} i \lambda_{hhh}
  + \frac{\partial \Psi}{\partial H}
  \frac{i}{s-M_H^2+i M_H \Gamma_H} i \lambda_{Hhh} \right) \nonumber \\
&&  \left(
   -4 + \frac{s-4m_N^2+A}{2 b} \ln\left|\frac{A+2 b}{A-2 b}\right|
  \right) ,
\end{eqnarray}
 where $\theta$ is the scattering angle in the center of mass frame.
The auxiliary functions appear above are defined as
\begin{eqnarray} 
I_{22}(s)
 &\equiv& 4 \frac{
   (A+2a)^2 -2(s+4m_N^2)A-s(A+m_N^2)-3m_N^2(s-4m_N^2)  }{A^2-4 b^2} , \\
J_{22}(s,m_h)
 &\equiv&  \frac{1}{A b}\left(
     2A(A+2a)-A(s+4m_N^2)+A^2-4a^2-(s-2m_N^2)(m_N^2-m_h^2) \right.\nonumber \\ 
 && \left. +3m_N^2(s-4m_N^2)
           \right) , \\
 A(s,m_h) &\equiv& -\frac{s}{2}+m_h^2 , \\
 b(s,m_N,m_h) &\equiv& \sqrt{\frac{s}{4}-m_h^2}\sqrt{\frac{s}{4}-m_N^2} .
\end{eqnarray}

\section{Thermal averaged annihilation cross section}

In partial wave expansion, the thermal averaged cross section 
is given by 
\begin{eqnarray}
\langle \sigma v \rangle 
&=& \frac{1}{m_N^2}\left.\left[
w(s) -\frac{3}{2}\left(2 w(s) -4 m_N^2 \frac{d w}{d s} \right)\frac{T}{m_N}
\right]\right|_{s=4m_N^2} \\
&=& 6 \left.\frac{d w}{d s}\right|_{s=4m_N^2}  \frac{T}{m_N}, 
\end{eqnarray}
with 
\begin{eqnarray}
 4 w(s) \equiv \int d{\rm LIPS} \sum|{\cal M}|^2 = \frac{1}{8 \pi}\sqrt{\frac{s-4 m^2_{\rm final}}{s}} 
 \int\frac{d \cos\theta}{2}
 \sum|{\cal M}|^2 ,
\end{eqnarray}
where $m_{\rm final}$ is the mass of final state particle.



\end{document}